\documentclass[twoside,english,aps,pra,superscriptaddress,longbibliography, notitlepage, twocolumn]{revtex4-1}
\usepackage[T2A,T1]{fontenc}
\usepackage[latin9]{inputenc}
\usepackage{color}
\usepackage{amsmath}
\usepackage{amssymb}
\usepackage{graphicx}

\makeatletter

\usepackage[colorlinks,citecolor=red,urlcolor=blue,bookmarks=false,hypertexnames=true]{hyperref} 

\makeatother

\usepackage{babel}
\begin{document}

\title{Interplay of hyperbolic plasmons and superconductivity}
\begin{abstract}
Hyperbolic plasmons are collective electron excitations in layered conductors. They are of relevance to a number of superconducting materials, including the cuprates and layered hyperbolic metamaterials  {[V. N. Smolyaninova, \textit{et al.}  Scientific Reports \textbf{6}, 34140 (2016)]}. This work studies   how the unusual dispersion of hyperbolic plasmons affects Cooped pairing. We use the Migdal-Eliashberg equations, which are solved both numerically and analytically using the Grabowski-Sham approximation, with consistent results.  We do not find evidence for plasmon-mediated pairing within a reasonable parameter range. However, it is shown  that the hyperbolic plasmons can significantly reduce the effects of Coulomb repulsion in the Cooper channel leading to an enhancement of the transition temperature originating from  other pairing mechanisms. In the model of a hyperbolic material composed of identical layers, we find this enhancement to be the strongest in the $d$-wave channel. We also discuss strategies for engineering an optimal hyperbolic plasmon background  for a further enhancement of superconductivity in both $s$-wave and $d$-wave channels.
\end{abstract}

\author{Andrey Grankin and Victor Galitski}

\affiliation{Joint Quantum Institute, University of Maryland, College Park, MD
20742, USA}
\affiliation{ Department of Physics, University of Maryland, College Park, MD 20742, USA}

\maketitle

\section{Introduction}

\begin{figure}
\begin{centering}
\includegraphics[scale=0.45]{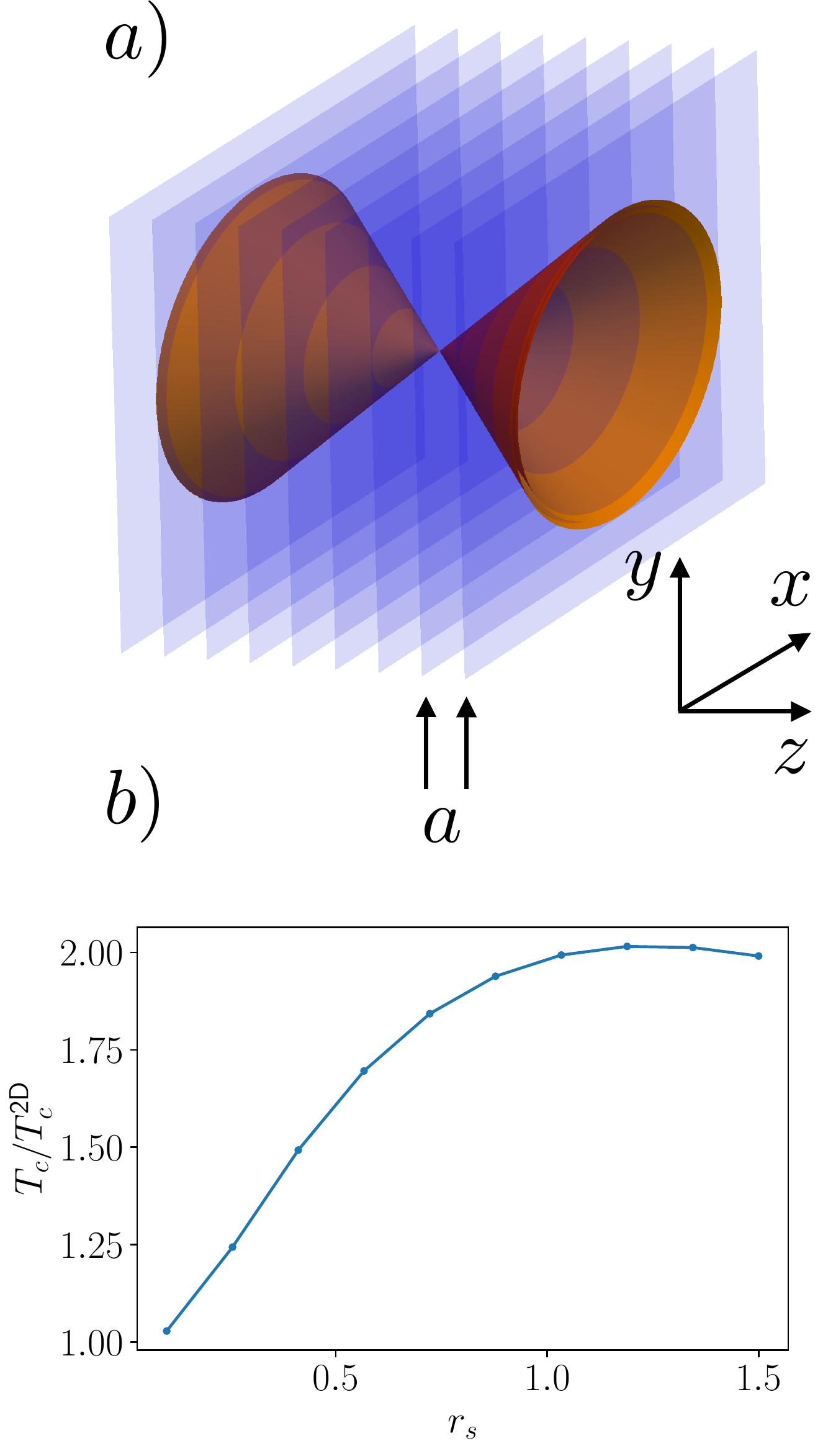}
\par\end{centering}
\caption{Superconductivity in a layered electron gas. a) Sketch of the setup:
 layered electron gas with inter-layer distance $a$.
Hyperbolic plasmon dispersion is shown in orange. b)~Enhancement
of superconductivity in a $d$-wave channel: coupling constant $\mu^{*}$ 
as function of the Wigner-Seitz parameter $r_s$. The ratio between the superconducting transition temperatures in the layered electron gas and the 2DEG. The additional pairing interaction
parameters are chosen as $\lambda=0.5$, $\omega_0=0.1E_{F}$. The
interlayer distance $a=q_{0}^{-1}$ corresponds to the inverse of
the Thomas-Fermi screening vector. }

\label{FIG0}
\end{figure}

Plasmons are collective excitations of conducting electrons \citep{BAS21}.
In strongly anisotropic conductors, the plasmon dispersion is hyperbolic
and can strongly modify the electromagnetic properties of such materials
\citep{SAJ14,HEQ88}. Some examples include the possibility of engineering
a negative refraction index and lensing \citep{PIB13} beyond the
diffraction limit \citep{SBOOK}. Hyperbolic materials were also proposed to enable
long-range dipole-dipole interactions \citep{CJ17}, which is potentially
useful for efficient quantum information processing and transfer. 

Hyperbolic plasmons in layered metals \citep{DQ82, JD87} have the density of states, which
is linear in energy,  as was theoretically shown in \citep{MbK93}.
As a result,  such plasmons are expected to strongly influence superconducting properties
of hyperbolic materials.  The effect of layering on superconductivity has been theoretically studied in a simplified
way in Refs.~\citep{KM88,BMK03}, where it was argued  that the layering
can increase the transition temperature. 

The possibility of superconductivity mediated by two- and three-dimensional bulk plasmons
was first theoretically considered in the early works by Takada~\citep{T78,T92} within the 
so-called Kirzhnits-Maksimov-Khomskii (KMK) \cite{KMK73} approximation of the Migdal-Eliashberg equations.  
In particular, it was suggested that pairing can occur at low charge carrier densities. 
However, it was later realized (and we confirm this conclusion here) that the KMK approximation is not reliable.
In particular, it  underestimates the  transition temperature \citep{SS83,KA80} in the case of the phonon-induced superconductivity and also fails to 
correctly describe the plasmonic mechanism of Cooper pairing \cite{RS83}.

The proper treatment of the plasmon-induced and plasmon-assisted
pairing requires the use of full Migdal-Eliashberg equations \citep{GS84}, which is the focus of this paper, where we focus on layered materials. This research is partially motivated by recent experimental works \citep{SJZ16,SYZ14}, which demonstrated a strong enhancement of superconductivity in meta-materials with epsilon-near-zero (ENZ) properties \citep{LE17, SBOOK}. In particular, this  has been  achieved in a layered structure composed of conducting aluminium layers separated by thin dielectric layers of aluminium oxide. Due to the strong anisotropy of such materials, the plasmon spectrum is hyperbolic with the dielectric constant approaching zero along
a set of cones in the phase space, as schematically shown in Fig.~\ref{FIG0}~(a).
Based on the KMK arguments in \citep{SJZ16,SYZ14} the authors directly attribute the observed increased critical temperature to the plasmon-induced mechanism of pairing. 
In the current work by solving the Migdal-Eliashberg equations we rule out the possibility of the superconductivity induced by hyperbolic plasmons only. 
Therefore the enhancement of the critical temperature  in \citep{SJZ16,SYZ14} must be explained by other mechanisms e.g. hyperbolic phonons, which will be considered elsewhere. 
However, we find here that hyperbolic plasmons in combination with an additional intrinsic attractive interaction can lead to an enhancement of pairing. 

In this work we  perform a systematic study of the effect of hyperbolic
plasmons on superconductivity.  We consider both pairing due
to plasmon mechanism and a hyperbolic plasmon-assisted pairing on top of another
intrinsic pairing mechanism, e.g., due to phonons or magnons. Within the Random Phase Approximation (RPA) we numerically 
solve the complete set of Migdal-Eliashberg equations
\citep{M20}, and as shown in Fig.~\ref{FIG0}~(b), find a significant enhancement of superconductivity
in the $d$-wave channel and in the presence of additional attractive
pairing mechanism in the corresponding channel. The latter may be of relevance to  high-temperature superconductivity in cuprates \cite{TK00}. In order to get a qualitative
insight  into our numerical solution, we employ the Grabowski-Sham approximation \citep{GS84} and consider the band-averaged electron-electron interaction. We find that the
effect of hyperbolic plasmons on the transition temperature is two-fold.
On the one hand, they screen the Coulomb repulsion more efficiently compared
to the conventional two-dimensional plasmons. On the other, they
have a larger effective screening energy range. These two effects
are found to compete in how they impact the superconducting transition temperature. We
also consider the possibility of engineering \citep{L64} a plasmon
structure of layered materials to optimize the enhancement effect on superconductivity.
One proposed proof-of-principle scenario involves a layered structure, where one of the layers  is distinct from the other identical layers. We demonstrate a possible significant enhancement
of the superconducting transition temperature in such a configuration in both $s$-wave and $d$-wave channels.

This paper is structured as follows:  Sec.~\ref{sec:Formulation}
introduces the Migdal-Eliashberg formalism that takes into account retardation effects. In Sec.~\ref{sec:Numerics}, we solve
the Migdal-Eliashberg equations numerically for pure Coulomb interaction in the layered electron
gas. To elucidate the numerical results, in Sec.~\ref{sec:Discussion}
we introduce several approximation schemes that allow us to provide an intuitive picture of the hyperbolic plasmon-mediated pairing. 
In Sec.~\ref{sec:Effect-on-other}, we consider the effect of hyperbolic plasmons on other pairing mechanisms and demonstrate a possible enhancement of pairing the $d$-wave channel. 
 Sec.~\ref{subsec:Engineering-strong-plasmon} provides an example of a setup for the further enhancement of pairing by engineering the plasmon
dispersion in a hyperbolic/layered structure. 

\section{Formulation\label{sec:Formulation}}

We consider a layered electron gas with the inter-layer spacing $a$ (see, Fig.~\ref{FIG0}) \cite{DQ82}, which can be  either a meta-material structure~\cite{SBOOK} or a layered compound (e.g., a cuprate material). We describe electron properties of each layer by a Coulomb-interacting 2-dimensional electron  gas (2DEG) model. The full Hamiltonian reads:

\begin{align}
\hat{H}_{\text{full}} & =\hat{H}_{0}+\hat{H}_{\text{int}}, \label{eq:H_full}\\
\hat{H}_{0} & =\sum_{i,{\bf k},\sigma}\xi_{k}\hat{\psi}_{{\bf k},\sigma}^{\left(i\right)\dagger}\hat{\psi}_{{\bf k},\sigma}^{\left(i\right)},\\
\hat{H}_{\text{int}} & =\frac{1}{2A}\sum_{i\leq j,{\bf k,k',q},\sigma,\sigma^{\prime}}\tilde{V}_{{\bf q}}^{\left(i,j\right)}\hat{\psi}_{{\bf k}+{\bf k},\sigma}^{(i)\dagger}\hat{\psi}_{{\bf k}^{\prime}-{\bf q},\sigma^{\prime}}^{(j)\dagger}\hat{\psi}_{{\bf k}^{\prime},\sigma^{\prime}}^{(j)}\hat{\psi}_{{\bf k},\sigma}^{(i)},
\end{align}
where $\hat{\psi}_{{\bf k},\sigma}^{\left(i\right)}$ denotes the
creation/annihilation operator of spin-$\sigma$ electron on $i$-th
layer. $\xi_{k}=k^{2}/2m-E_F$ where $m$ refers to the bare
electron mass,  $A$ is the surface area of each layer and $E_F$ is the Fermi energy. $\tilde{{\cal V}}_{{\bf q}}^{\left(i,j\right)}=2\pi e^{2}e^{-aq\left|i-j\right|}/\epsilon_{\infty}q$
denotes the bare Coulomb interaction of electrons on $i$-th and $j$-th
layers, $a$ denotes the interlayer distance and $\epsilon_{\infty}$
is the background dielectric constant and $e$ is the electron charge.
In the following ${\bf q}$ denotes the two-dimensional in-plane momentum.
In our model Eq.~\eqref{eq:H_full} we neglected the electron tunnelling between layers. In Fourier space with respect to the layer index the bare Coulomb
interaction is

\[
\tilde{{\cal V}}_{{\bf q},q_{z}}=\frac{2\pi e^{2}}{\epsilon_{\infty}q}\frac{\sinh aq}{\cosh aq-\cos q_{z}},
\]
where $q_{z}\in[-\pi,\pi]$ denotes the out-of-plane momentum. The
conventional RPA-renormalized  interaction reads:

\begin{equation}
{\cal V}_{{\bf q},q_{z}}\left(i\Omega_{m}\right)=\frac{1}{\tilde{{\cal V}}_{{\bf q},q_{z}}^{-1}+\frac{m}{\pi}\Pi_{{\bf q}}\left(i\Omega_{m}\right)}\label{eq:Vii}
\end{equation}
where $\Pi_{{\bf q}}$ is the  polarization operator of
the 2D electron gas normalized as $\Pi_{{\bf q}\rightarrow0}\left(0\right)=1$
and $\Omega_{m}=2\pi m/\beta$. 

\begin{equation}
{\cal V}_{{\bf q},q_{z}}\left(i\Omega_{m}\right)=\frac{2\pi e^{2}}{\epsilon_{\infty}}\frac{1}{q\frac{\cosh aq-\cos q_{z}}{\sinh aq}+q_{0}\Pi_{{\bf q}}\left(i\Omega_{m}\right)},
\end{equation}
where $q_{0}=2me^{2}/\epsilon_{\infty}$ denotes the Thomas-Fermi
screening length. The latter characterizes the momentum-scale below
which the interaction is efficiently screened. 

Within the RPA the  normal-state electron  Green's function has the following form:

\begin{align}
{\cal G}_{{\bf k}}\left(i\varepsilon_{n}\right) & \equiv-\int_{0}^{\beta}e^{i\varepsilon_{n}\tau}\left\langle \hat{\psi}_{{\bf k,\sigma}}^{\left(i\right)}\left(\tau\right)\hat{\psi}_{{\bf k},\sigma}^{\left(i\right)\dagger}\left(0\right)\right\rangle \nonumber \\
 & =\frac{1}{-\xi_{k}+i\varepsilon_{n}-\Sigma_{{\bf k}}\left(i\varepsilon_{n}\right)}\label{eq:Gq}
\end{align}
where $\varepsilon_{n}=\left(2n+1\right)\pi/\beta,n\in\mathbb{Z}$ and
$\beta$ denotes the inverse temperature. As we assume all layers
to be identical and neglect tunnelling between layers, the Green function
in Eq.~(\ref{eq:Gq}) is independent of the layer index. The normal state
self-energy denoted above as $\Sigma_{{\bf k}}$ obeys the Dyson equation:

\begin{align}
\Sigma_{{\bf k}}\left(i\varepsilon_{n}\right) & =-\frac{1}{\beta A}\sum_{{\bf k'},m}G_{{\bf k'}}\left(i\varepsilon_{m}\right){\cal V}_{{\bf k}-{\bf k'}}^{\left(i,i\right)}\left(i\varepsilon_{m}-i\varepsilon_{n}\right),\label{eq:Sigma}
\end{align}

where ${\cal V}_{{\bf q}}^{(i,j)}(i\Omega_{m})=(2\pi)^{-1}\int dq_{z}e^{iq_{z}(i-j)}{\cal V}_{{\bf q},q_{z}}(i\Omega_{m})$ denotes the Fourier transform of the interaction Eq.~\eqref{eq:Vii}. The superconducting anomalous self-energy on $i$-th layer obeys the following equation:
\begin{align}
\phi_{{\bf k}}\left(i\varepsilon_{n}\right) & =-\left(\beta A\right)^{-1}\sum_{{\bf k'},m}{\cal V}^{(i,i)}_{{\bf k}-{\bf k'}}\left(i\varepsilon_{n}-i\varepsilon_{m}\right)\nonumber \\
 & \times{\cal G}_{{\bf k'}}\left(i\varepsilon_{m}\right){\cal G}_{-{\bf k'}}\left(-i\varepsilon_{m}\right)\phi_{{\bf k'}}\left(i\varepsilon_{m}\right).\label{eq:phi_k}
\end{align}
We solve this equation numerically for both pure Coulomb interaction and it co-existing with other pairing mechanisms. We also consider simplified analytical
models for both cases, which help us develop an intuitive physical picture behind the numerical results. 

\subsection{Hypebolic plasmon dispersion}

\begin{figure}
\begin{centering}
\includegraphics[scale=0.3]{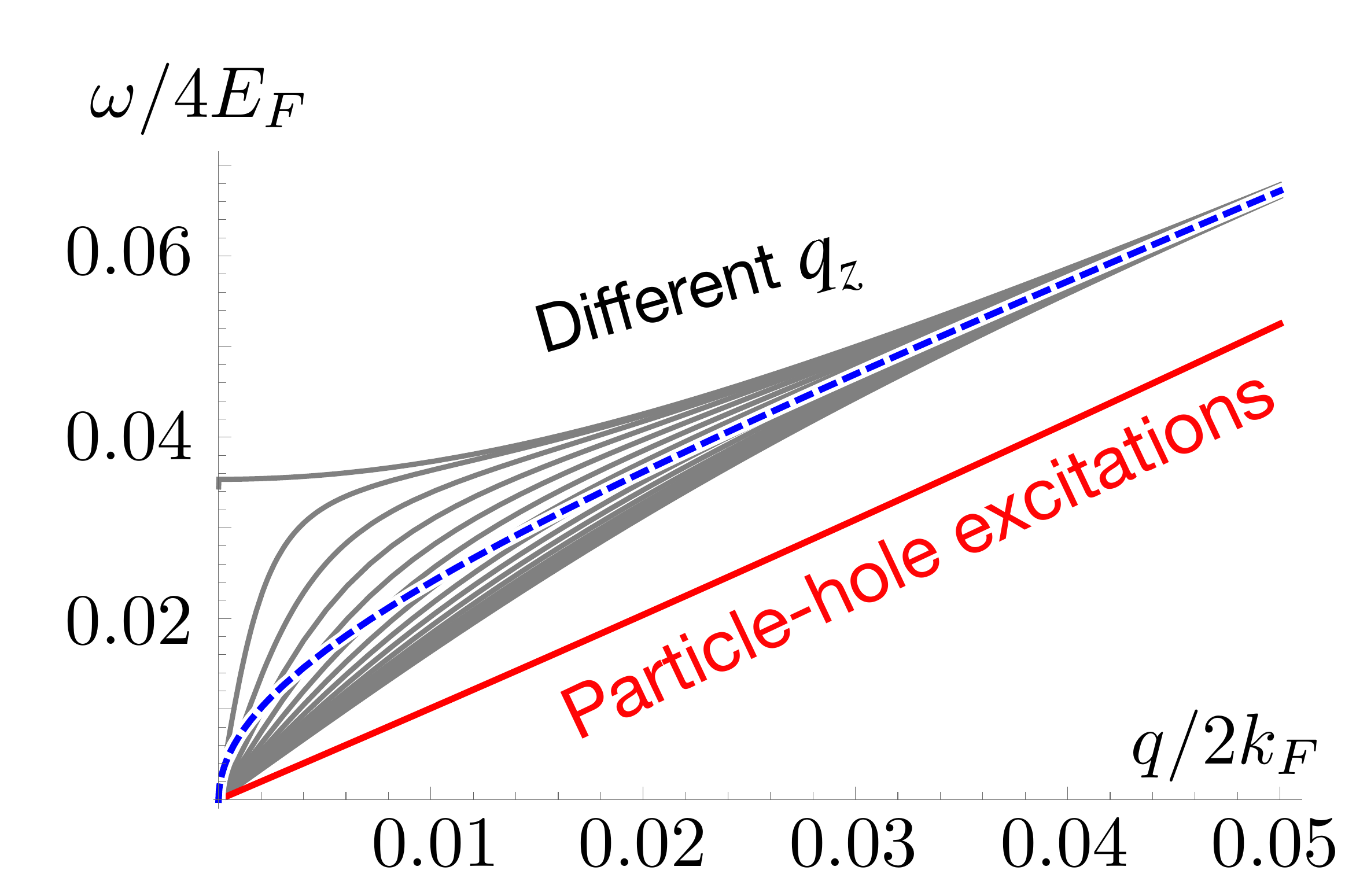}
\par\end{centering}
\caption{Layered plasmon spectrum for the interlayer spacing $a=1/q_{0}$. Different gray curves correspond
to different values of $q_{z}\in\left[0,\pi\right]$ . Blue dashed
line stands for the pure 2-dimensional plasmon spectrum. Red curve
corresponds to the particle-hole excitation spectrum of the electron gas.}

\label{Fig_plasmon}
\end{figure}

We now describe the plasmon structure of the layered electron gas
defined above. For that we perform the analytic continuation of the Coulomb propagator to real-frequency: 

\begin{equation}
\text{Im}V_{q,q_{z}}(\omega+i0^{+})=\frac{2\pi e^{2}}{\epsilon_{\infty}}\text{Im}\frac{1}{q\frac{\cosh aq-\cos q_{z}}{\sinh aq}+q_{0}\Pi_{q}\left(\omega\right)},
\end{equation}
Its poles correspond to the collective excitations of the layered electron
gas (plasmons). We readily find the long-wavelength plasmon excitation
dispersion for $q_{z}\neq0$:

\begin{equation}
\varepsilon_{q}=v_F\sqrt{1 + \frac{a q_0}{2 \sin^2(q_z/2)}}q,\label{eq:w_q}
\end{equation}
where $v_F=k_F/m$ is the Fermi velocity. For $q_{z}=0$ the spectrum is gapped and the dispersion is $\varepsilon_{q\rightarrow0}\approx v_F\sqrt{q_{0}/a}$.
For any non-zero $q_{z}$ the fixed-frequency curves form cones in
momentum space as schematically shown in Fig.~\ref{FIG0}~(a). The
full plasmon spectrum is shown in Fig.~\ref{Fig_plasmon}. We note several features
of the  spectrum Eq.~(\ref{eq:w_q}). First, it has a finite density
of states at low energies. Second, it extends to higher frequencies
compared to the pure two-dimensional plasmons. 

\subsection{Angular averaging. Migdal-Eliashberg equations.}

To solve  Eq.~\eqref{eq:phi_k}, we first project the gap function onto a particular angular momentum eigenstate  $\phi_{{\bf k}}\propto e^{il\vartheta}+c.c.$,
where $\vartheta$ is the angular coordinate. For the sake of convenience below we express all
energy and momentum variables in $4E_{F}$ and $2k_{F}$ respectively:
$k\rightarrow2k_{F}k$, $a\rightarrow a/2k_{F}$, $\beta\rightarrow\beta/4E_{F}$.  In these units the normal-state Green function has the following form ${\cal G}_{k}^{-1}\left(i\varepsilon_{m}\right)=i\varepsilon_{m}-k^{2}+1/4-\Sigma_{\bf{k}}$.

Let us now explicitly consider the gap equation  Eq.~\eqref{eq:phi_k} in the $l$-wave channel. By transforming summations into integrals we readily find:

\begin{align}
\phi_{{\bf k}}^{(l)}\left(i\varepsilon_{n}\right)= & -T\sum_{n}\int k'dk'\int_{0}^{2\pi}\frac{d\vartheta}{2\pi}\int_{-\pi}^{\pi}\frac{dk_{z}}{2\pi}\nonumber \\
 & \times\frac{q_{0}}{u\frac{\cosh au-\cos k_{z}}{\sinh au}+q_{0}\Pi_{u}}\frac{\phi_{{\bf k'}}^{(l)}\left(i\varepsilon_{m}\right)}{\left|{\cal G}_{k'}^{-1}\left(i\varepsilon_{m}\right)\right|^{2}},
\end{align}
where $u\equiv\left|{\bf k}-{\bf k'}\right|=\sqrt{k^2+k'^2-2kk' \cos(\vartheta)}$. As we see from the  expression
the  result depends only on two variables: $q_{0}$ and $a$. Now assuming a
particular angular momentum $l$ of the gap we can perform the
angular integral:

\begin{align}
 & \phi_{k}^{(l)}\left(i\varepsilon_{n}\right)=-T\sum_{m}\int k'dk'{\cal V}_{k,k'}^{(l)}\left(i\varepsilon_{n}-i\varepsilon_{m}\right)\frac{\phi_{{\bf k'}}^{(l)}\left(i\varepsilon_{m}\right)}{\left|{\cal G}_{k'}^{-1}\left(i\varepsilon_{m}\right)\right|^{2}},\label{eq:phi}
\end{align}
where the angular-averaged interaction is defined as:

\begin{align}
 & {\cal V}_{k,k'}^{(l)}\left(i\Omega_{n}\right)=  \int\frac{d\vartheta}{2\pi}\int_{-\pi}^{\pi}\frac{dk_{z}}{2\pi}\frac{q_{0}\cos(\vartheta l)}{u\frac{\cosh au-\cos q_{z}}{\sinh au}+q_{0}\Pi_{u}} \nonumber \\
  & =\int\frac{d\vartheta}{2\pi}\frac{q_{0}\sinh (au) \cos(\vartheta l)}{\sqrt{\left(u\cosh au+q_{0}\sinh au\Pi_{u}\left(i\Omega_{n}\right)\right)^{2}-u^{2}}\label{eq:Vkkp}}
\end{align}
We note that the propagator for the nearest-neighbor layers can also be found analytically. Analogously,
we find the corresponding equation for the normal-state self-energy:

\begin{align}
\Sigma_{k}\left(i\varepsilon_{n}\right) & =-T\sum_{m}\int k'dk'{\cal V}_{k,k'}^{(l=0)}\left(i\varepsilon_{m}-i\varepsilon_{n}\right){\cal G}_{k'}\left(i\varepsilon_{m}\right).\label{eq:Sigma-1}
\end{align}

We now define the conventional 2-dimensional Wigner-Seitz parameter
$2\pi a_{\text{Bohr}}^{2}r_{s}^{2}n=1$, where $a_{\text{Bohr}}$ is
the Bohr radius in dielectric with the polarizability $\epsilon_{\infty}$.
$r_{s}$ can be expressed via the dimensionless Thomas-Fermi vector
as $r_{s}=\sqrt{2}q_{0}$. In the following section we solve Eqs.~(\ref{eq:phi},
\ref{eq:Sigma-1}) numerically.

\section{Numerical solution of Migdal-Eliashberg equations\label{sec:Numerics}}

In this section we provide details of numerical solution of Migdal-Eliashberg
equations  Eqs.~(\ref{eq:phi},
\ref{eq:Sigma-1}). In Sec.~\ref{subsec:Normal-state-self-energy}
we discuss the calculation of the normal-state self energy. We then
numerically solve the gap equation for $s$- and $d$-wave pairing channels. 
We find that the plasmon induced pairing is impossible in the reasonable parameter range.

\subsection{Normal state self-energy\label{subsec:Normal-state-self-energy}}

The self-consistent solution of the gap equation in case of Coulomb interaction requires caution as  the interaction kernel in  Eqs.~(\ref{eq:phi},
\ref{eq:Sigma-1}) 
has  a logarithmic singularity for $k=k'$ \citep{RS83}. We therefore introduce a momentum
grid which has density of momentum points with the smallest $\Delta k\propto10^{-3}k_{F}$.
 We first solve Eq.~\eqref{eq:Sigma-1} iteratively. As we explain in more detail below,  it is convenient to make the
conventional variable change \citep{RS83}:

\begin{align}
\chi_{k}\left(i\varepsilon_{n}\right)= & \frac{\Sigma_{k}\left(i\varepsilon_{n}\right)+\Sigma_{k}\left(-i\varepsilon_{n}\right)}{2},\label{eq:chi}\\
Z_{k}\left(i\varepsilon_{n}\right)= & 1+i\frac{\Sigma_{k}\left(i\varepsilon_{n}\right)-\Sigma_{k}\left(-i\varepsilon_{n}\right)}{2\varepsilon_{n}},\label{eq:Zk}
\end{align}
where from Eq.~\eqref{eq:Sigma-1}  $Z_{k}\left(i\varepsilon_{n}\right)$ is found to obey: 

\begin{align}
 & \varepsilon_{n}\left(Z_{k}\left(i\varepsilon_{n}\right)-1\right)=\nonumber \\
 & T\int k'dk'\sum_{m}\frac{\varepsilon_{m}Z_{k'}\left(i\varepsilon_{m}\right)}{\left|{\cal G}_{k'}^{-1}\left(i\varepsilon_{m}\right)\right|^{2}}{\cal V}_{k,k'}^{(l=0)}\left(i\varepsilon_{m}-i\varepsilon_{n}\right),\label{eq:Eq_Zk}
\end{align}
The even-frequency part of the normal-state self-energy  $\chi_{k}\left(i\varepsilon_{n}\right)$ function has static and dynamical contributions:

\begin{align}
\chi_{k}\left(i\varepsilon_{n}\right) & =\chi_{k}^{\left(0\right)}+\chi_{k}^{\left(1\right)}\left(i\varepsilon_{n}\right)\\
\chi_{k}^{\left(0\right)} & =-T\int k'dk'\sum_{m}\frac{\xi_{k'}+\chi_{k'}\left(i\varepsilon_{m}\right)}{\left|{\cal G}_{k'}^{-1}\left(i\varepsilon_{m}\right)\right|^{2}}{\cal \tilde V}_{k,k'},\\
\chi_{k}^{\left(1\right)}\left(i\varepsilon_{n}\right) & =-T\int k'dk'\sum_{m}\frac{\xi_{k'}+\chi_{k'}\left(i\varepsilon_{m}\right)}{\left|{\cal G}_{k'}^{-1}\left(i\varepsilon_{m}\right)\right|^{2}}\delta{\cal V}_{k,k'}^{(l=0)}\left(i\varepsilon_{m}-i\varepsilon_{n}\right)
\end{align}
where ${\cal \tilde V}_{k,k'}$ refers to the bare unrenormalized
Coulomb interaction and $\delta{\cal V}_{k,k'}^{(l)}\equiv{\cal V}_{k,k'}^{(l=0)}-{\cal \tilde V}_{k,k'}$.
The advantage of this representation is that $\chi_{k}^{\left(0\right)}$
is frequency-independent while $\chi_{k}^{\left(1\right)}$ has finite
frequency range and can be solved by introducing a frequency cut-off.
Equations for $\chi$ and $Z$ can be solved iteratively starting
with some initial value \citep{RS83}. 

\subsection{Numerical procedure\label{subsec:Solution-of-gap}}

For solution of Eqs.~\eqref{eq:phi}, we note that the gap function is frequency-dependent and does not vanish at large frequencies:

\begin{equation}
\phi_{k}^{\left(i,j\right)}\left(\pm i\infty\right)=-T\sum_{m}\int k'dk' \frac{\phi_{k'}^{\left(i,j\right)}\left(i\varepsilon_{m}\right)}{\left|{\cal G}_{k'}^{-1}\left(i\varepsilon_{m}\right)\right|^{2}} {\cal \tilde V}_{k,k'}.\label{eq:phi-2}
\end{equation}
This implies that we cannot impose a simple frequency cutoff in the integral equation. Here instead we introduce a cut-off, $\epsilon_*$, above which the gap function is assumed constant and reaches its infinite-frequency value $\phi_{k'}^{(i,j)}(i|\varepsilon_{n}| > \varepsilon_{\epsilon_*})=\phi_{k'}^{(i,j)}\left(i\infty\right)$.
The value of $\epsilon_*$ is chosen to ensure the convergence of the iterative procedure and by the condition that a further increase of  $\epsilon_*$ does not affect the transition temperature withing desired accuracy.

{
With the above in mind we now describe the iterative numerical procedure for solving Eqs.~(\ref{eq:phi},
\ref{eq:Sigma-1}). 
We start with the iterative procedure with the Green function for a free Fermi gas; i.e. $\Sigma^{(0)}_k=0$. 
Using the decomposition into odd- and even-frequency parts described in Sec.~\ref{subsec:Normal-state-self-energy}, we numerically apply the integral operator on the right-hand side of Eq.~\eqref{eq:Sigma-1} $n$ times and find  $\Sigma^{(n)}_k$. 
It is typically sufficient to perform approximately 10 iterative steps until the difference $|\Sigma^{(n)}_k-\Sigma^{(n-1)}_k|\ll |\Sigma^{(n)}|$ is negligible and the iteration scheme converges. Once normal-state self energy is obtained we can solve the gap equation Eq.~\eqref{eq:phi}. 
A solution of this equation determines the critical temperature $T_c$. 
Equivalently, at the transition temperature the discretized integral operator on the right-hand side of Eq.~\eqref{eq:phi} has eigenvalue 1. 

We find the transition point by changing temperature as a parameter of the integral kernel.}

\subsection{Numerical results\label{subsec:Numerical_results}}

We now discuss the results of the numerical solution of the gap equation Eq.~\eqref{eq:phi} according to the procedure outlined above. 
In addition to the pure plasmon-induced superconductivity, we also consider the case when some additional attraction between electrons is present. For the latter case here we only describe the final result leaving most of the details to Sec.~\ref{sec:Effect-on-other}.

 At the electron densities corresponding to the regime of validity of the RPA, we find no possible plasmon-induced pairing in both layered and in the conventional 2-dimensional electron gas regime. While in agreement with the earlier studies \citep{T78},\citep{RS83} the solution of Eliashberg equations exists for $r_s>r^*_s$, where $r^*_s\approx2$, which is beyond the applicability of the RPA. In this case, it is important to keep vertex corrections which may qualitatively change the result. 
 Extrapolating the results outside the RPA validity range we find that the layering is decreasing the transition temperature for $r_s>r^*_s$. The same behavior occurs in the $d$-wave channel as well.
In conclusion, within the RPA we do not find a regime where plasmon-induced pairing is possible.  
 
The situation is, however, different  if we add an additional attractive interaction in the $d$-wave channel. 
More precisely, we consider the following modification of the Coulomb interaction Eq.~\eqref{eq:Vkkp}:
\begin{align}
{\cal V}_{k,k'}^{(l)}(i\Omega_{n})\rightarrow{\cal V}_{k,k'}^{(l)}(i\Omega_{n})-\lambda\delta_{l,2}\frac{\omega_0^{2}}{\omega_0^{2}+\Omega_{n}^{2}}\label{eq:Vkkmod},
\end{align}
where $\lambda$ is some effective interaction strength and $\omega_0$ is the characteristic frequency. 
We note that the additional interaction in Eq.~\eqref{eq:Vkkmod} can be induced by a single-frequency bosonic mode.
As shown in Fig.~\ref{FIG0}~(c), we find an enhancement of the transition temperature in a layered electron gas compared to the conventional 2-dimensional electron gas. We note that depending on the parameters the observed enhancement can be quite significant. 
In particular for our choice of parameters in Fig.~\ref{FIG0}~(c) the critical temperature is doubled at $r_s=1$.
In the next section, we study a simplified model which, as we show, can explain all the features of the transition temperature.

\section{Toy model of plasmon-mediated pairing\label{sec:Discussion}}

\begin{figure}
\begin{centering}
\includegraphics[scale=0.25]{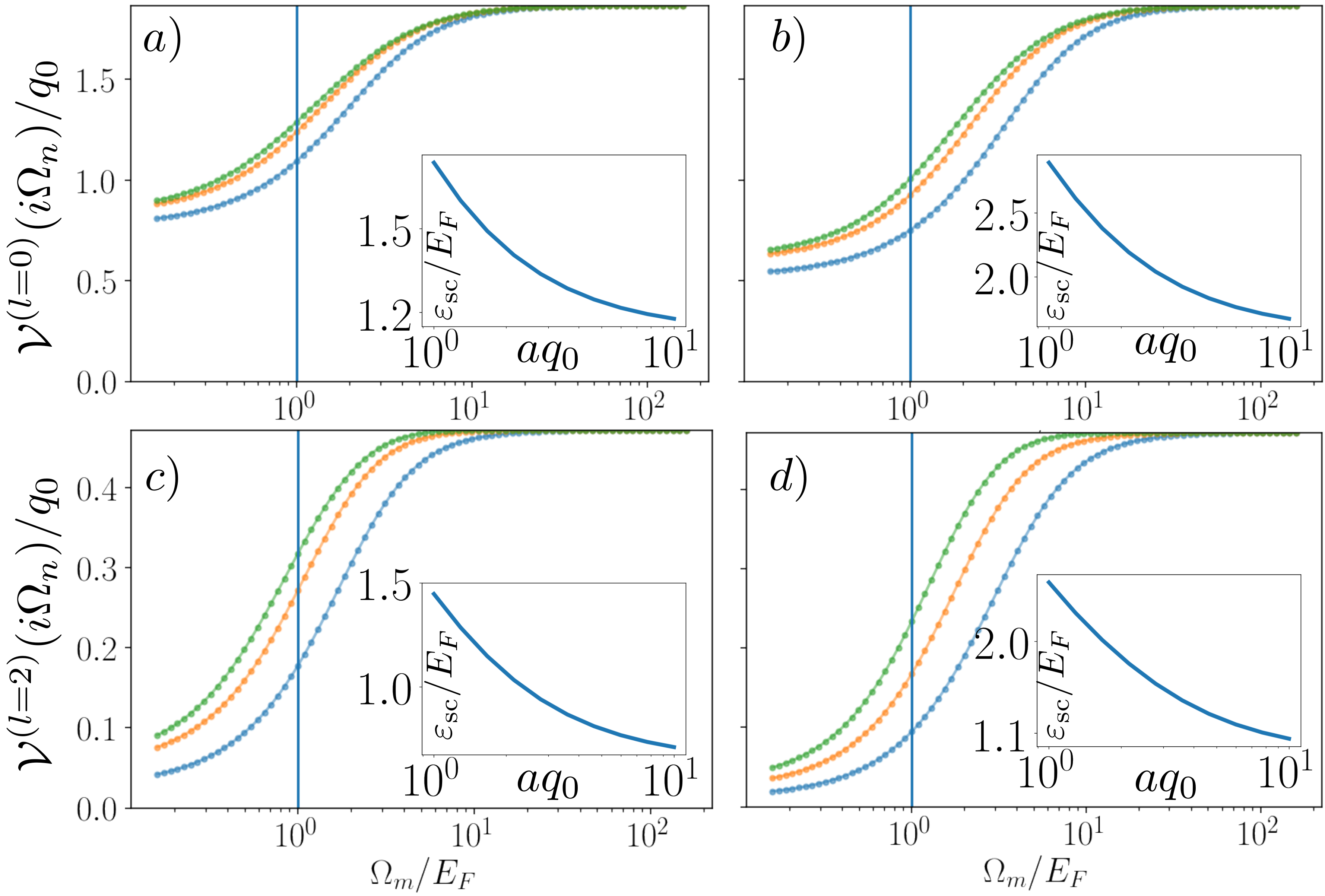} 
\par\end{centering}
\caption{Effective interaction ${\cal V}(i\Omega_{n})$ Eq.~(\ref{eq:V1d1})
for $aq_{0}=1$ (blue), $aq_{0}\approx3.2$ (orange) and $aq_{0}=10$
(green). $s$-wave channel with electron density is a) $r_{s}=0.8$,
b) $r_{s}=1.5$. $d$-wave channel with electron density is c) $r_{s}=0.8$,
d) $r_{s}=1.5$. Insets shows the effective screening energy
$\varepsilon_{\text{sc}}$ as function of the inter-layer distance $a$.
In both figures the blue vertical line corresponds to the Fermi energy. }

\label{Fig3} 
\end{figure}

In this section, we discuss a toy model that reproduces the behavior
of $T_{c}$ shown discussed above. 
The aim of this section is solely to develop an intuitive picture of the plasmon-mediated Cooper pairing.
For that, we use a simplified
model with no renormalization of the normal-state Green's function.
We start with Eq.~\eqref{eq:phi}:

\begin{align}
 & \phi_{k}^{(l)}\left(i\varepsilon_{n}\right)=\nonumber \\
 & -T\sum_{m}\int k'dk'{\cal V}_{k,k'}^{(l)}\left(i\varepsilon_{m}-i\varepsilon_{n}\right)\frac{\phi_{k'}^{(l)}\left(i\varepsilon_{m}\right)}{\varepsilon_{m}^{2}+\xi_{k'}^{2}}.\label{eq:phi-1}
\end{align}
To get an analytical insight we now reduce the complete 2-dimensional
Migdal-Eliashberg equation (\ref{eq:phi-1}) to an effective 1-D form.
Due to the long-range nature of Coulomb interaction, we cannot perform
the conventional Fermi-surface averaging of the interaction. Indeed,
at high Matsubara frequencies we find:

\[
{\cal V}_{k_{F},k_{F}}^{(l)}\left(i\Omega_{n}\right)\propto\int_{0}^{2\pi}d\vartheta\frac{\cos(\vartheta l)}{\sqrt{1+\cos\vartheta}}.
\]
The right-hand side of this integral logarithmically diverges at $\vartheta\sim\pi$
\citep{RS83}. This divergence is present for all frequencies except
for the zeroth one. to circumvent this technical issue, we resort to the heuristic Grabowski-Sham approximation
introduced in \citep{GS84}. Following their approach (with a slight modification that produces  more accurate results in comparison with the brute-force numerical solution of the original integral equation), we replace the
momentum-dependent interaction with:

\begin{equation}
{\cal V}^{(l)}\left(i\Omega_{n}\right)\equiv\frac{\int kdk\theta\left(k_{\text{c}}-k\right){\cal V}_{k,k_{F}}^{(l)}\left(i\Omega_{n}\right)}{\int kdk\theta\left(k_{\text{c}}-k\right)},\label{eq:V1d1}
\end{equation}
where $\theta$ is the Heaviside theta-function and $k_{c}$ is the cut-off momentum chosen according to the bandwidth
of the electron gas. Overall, we can conveniently parameterize \citep{GS84}
the interaction as follows: 
\begin{equation}
{\cal V}^{(l)}\left(i\varepsilon_{m}-i\varepsilon_{n}\right)\equiv\mu\left(1-\sigma v_{m,n}\right)\label{eq:Vtildetilde}
\end{equation}
with $v_{m,n\rightarrow\infty}\rightarrow0,v_{0,0}=1$ and $\sigma$
being a dimensionless  coefficient, characterizing screening strength. A typical example of this function is shown in Fig.~\ref{Fig3}. Here we observe  a suppression of the Coulomb interaction at Matsubara frequencies below some
characteristic screening energy scale $\varepsilon_{\text{sc}}$. The latter
can be heuristically inferred from the curves Fig.~\ref{Fig3} by
fitting $v_{m,n}$ with e.g. Lorentz curve (the result is shown in insets). 

Since we now ignore the momentum dependence of the interaction potential
the gap equation simplifies to:

\begin{align}
  \phi^{(l)}\left(i\varepsilon_{n}\right)=
  -T\sum_{m}{\cal V}^{(l)}\left(i\varepsilon_{m}-i\varepsilon_{n}\right)\int \frac{\phi^{(l)}\left(i\varepsilon_{m}\right) d\xi_{k'}}{\varepsilon_{m}^{2}+\xi_{k'}^{2}}.\label{eq:Eq_GS}
\end{align}

\begin{figure}
\begin{centering}
\includegraphics[scale=0.26]{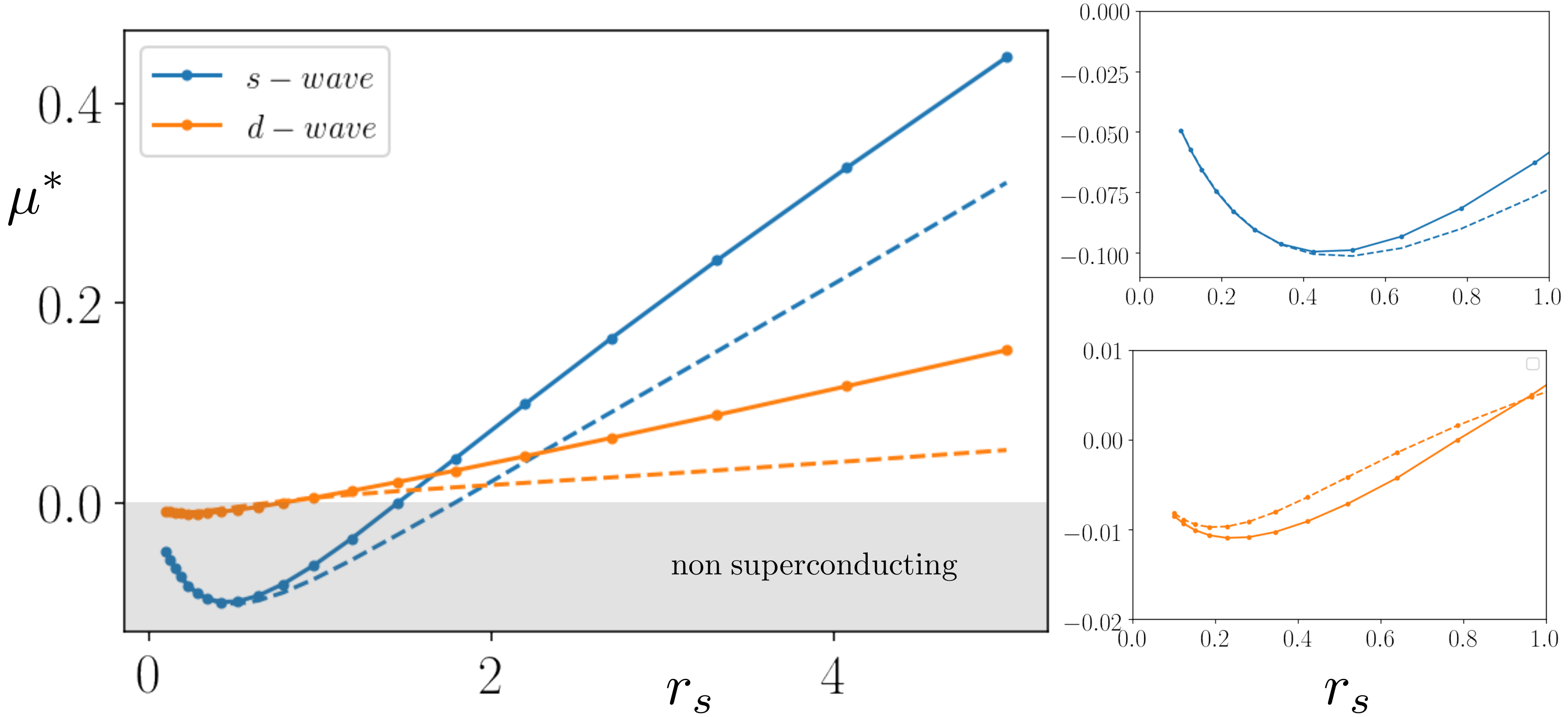} 
\par\end{centering}
\caption{Plasmon-induced coupling constant $\mu^{*}$
as function of the electron gas density $r_{s}$ for $s$-and $d$-wave
pairing channels. Solid line corresponds to an infinite inter-layer
distance $a\rightarrow\infty$, dashed corresponds to $aq_{0}=1$.
Right panel: detailed zoom on the electron density density range $r_{s}\in[0,1]$.}

\label{FIG_RGS} 
\end{figure}

Eq.~\eqref{eq:Vtildetilde} is one-dimensional and therefore it is straightforward to solve numerically. It is however more practical to use the
pseudo-potential method \cite{GS84}. It allows us to define the coupling strength even in the case when there is no superconductivity: $\mu^{*}<0$. 
In the latter case the quantity  $-\mu^{*}$ is typically referred to as the Coulomb pseudopotential.
The idea behind the pseudopotential method is to replace the long-range (in frequency) interaction
in Eq.~\eqref{eq:Eq_GS} with a short-range one. For the latter, the
critical temperature can be found trivially as discussed in the SM.
The superconducting coupling constant obtained from the approximation
Eq.~\eqref{eq:Vtildetilde} is shown in Fig.~\ref{FIG_RGS}. We
observe that in the regime when $\mu^{*}$ is positive the layering
is typically detrimental for the pairing in $s$- and $d$-wave channels.
The only regime where layering is useful is in $d$-wave channel at
$r_{s}\approx1$. 
\subsection{Analytical estimate of the transition temperature}
In order to explain the properties of the plasmon-mediated Cooper pairing it is instructive
to consider the exactly solvable model. More precisely, we assume
a separable square-well model for the interaction: $v_{m,n}\approx\theta(\varepsilon_{\text{sc}}^{2}-\varepsilon_{m}^{2})\theta(\varepsilon_{\text{sc}}^{2}-\varepsilon_{n}^{2})$,
where $\theta$ is the Heaviside theta-function. By inserting this
ansatz to Eq.~\eqref{eq:Eq_GS} in the limit $\varepsilon_{\text{sc}}\ll E_{F}$
the transition temperature is found to be (see Appendix~\ref{sec:Analytical-solutions}):

\begin{equation}
T_{c}=1.134 E_F e^{-1/\mu^*}\label{eq:Tc}
\end{equation}
with $\mu^{*}=-\log(E_F/\varepsilon_{\text{sc}})+\mu\left(\sigma-1/(1+\mu\log\frac{E_{F}}{\varepsilon_{\text{sc}}})\right)$.
The condition on existence of the superconductivity is given by $\mu^{*}>0$.
We also see  the suppression of the overall Coulomb repulsion term by the logarithmic prefactor. From Eq.~\eqref{eq:Tc} we find the optimal value $\varepsilon_{\text{sc}}=e^{\left(\sigma-2\right)/\mu\sigma}E_{F}$
at which the critical temperature is maximal. At larger values the
critical temperature is decreasing becoming exactly zero at $\varepsilon_{\text{sc}}=e^{\left(\sigma-1\right)/\mu\sigma}E_{F}$. 

With the above in mind we now analyze the properties of the effective
interaction ${\cal V}\left(i\Omega_{n}\right)$ shown in Fig.~\ref{Fig3}.
The effect of layering is found to be two-fold: first it increases the screening strength, $\sigma$,
and second, it increases the corresponding screening energy range $\varepsilon_{\text{sc}}$. As clear from Eq.~(\ref{eq:Tc}), the two effects discussed above contribute to $T_{c}$ in the opposite
way. Besides the negative effect is usually dominating. The only regime
when the pairing is enhanced is found in the $d$-wave channel at
smaller values $r_{s}<1$. This can be explained by the fact that
the effective interaction has the lowest screening energy scale $\varepsilon_{\text{sc}}$ and
the highest relative screening strength $\sigma$. We note that although $\mu^{*}$ is
enhanced by layering, the Cooper pairing itself is not possible due
to the overall sign of $\mu^{*}<0$. It  therefore appears impossible to achieve  a pure plasmon-induced superconductivity or an  enhancement thereof by hyperbolic plasmons (unless we consider high $r_s$ values, where no reliable theoretical methods exist).

In conclusion of this section, we reiterate that  the renormalization of the Coulomb pseudopotential by hyperbolic plasmons  contain two competing factors: the enhancement of the plasmon-induced screening parameterized by $\sigma$ and the increase of the effective screening energy range, $\varepsilon_{\text{sc}}$. In the case of  $d$-wave superconductivity, the microscopic details of the interaction favor the increase of the coupling constant $\mu^{*}$ thereby decreasing
Coulomb repulsion. In the next section we show that this effect can be beneficial in case when superconductivity is induced by other mechanisms.

\section{Plasmon-assisted pairing\label{sec:Effect-on-other}}
In this section we study how hyperbolic plasmons in a layered material affect other pairing mechanisms. We also provide an example of plasmon-engineering, where superconductivity is enhanced more efficiently. 

\subsection{Additional attractive interaction}
We now consider the Migdal-Eliashberg equations in the presence of additional attractive interaction e.g. phonon- or magnon-induced. A physical origin of the additional attraction is not important for the model, we study, where on top of the Coulomb repulsion we add another pairing interaction in the $l$-wave angular momentum channel as follows [c.f., Eq.~\eqref{eq:Vkkmod}]: 
$$
{\cal V}_{k,k'}^{(l)}(i\Omega_{n})\rightarrow{\cal V}_{k,k'}^{(l)}(i\Omega_{n})-\lambda\frac{\omega_0^{2}}{\omega_0^{2}+\Omega_{n}^{2}}.
$$

 The numerical result of the solution of Eqs.~\eqref{eq:phi} is shown in Fig.~\ref{FIG0}~(c) for the $d$-wave channel. Below we provide an explanation of the enhancement based on the toy model.

Analytically the transition temperature can be
estimated using the same formalism as in Sec.~\ref{sec:Discussion}.

\begin{equation}
T_{c}=1.134E_{F}e^{-1/\tilde{\mu}^{*}},\label{eq:TcLambda}
\end{equation}
where $\tilde{\mu}^{*}=\lambda_{E_{F}}+\mu^{*}$, where $\mu^{*}$ is the coupling constant due to plasmons only and the from the other pairing mechanism gives rise to the effective coupling $\lambda_{E_{F}}\equiv\lambda/\left[1+\lambda\log\left(E_{F}/\omega_0\right)\right]$
 renormalized to the Fermi energy scale. Provided that $\mu^*$ is approximately the same as we computed in the previous section we can expect the overall energy scale in Fig.~\ref{FIG_RGS} to be
lifted by the amount of this renormalized coupling $\lambda_{E_F}$. This leads to the enhancement of the $d$-wave pairing. 

\begin{figure}
\begin{centering}
\includegraphics[scale=0.3]{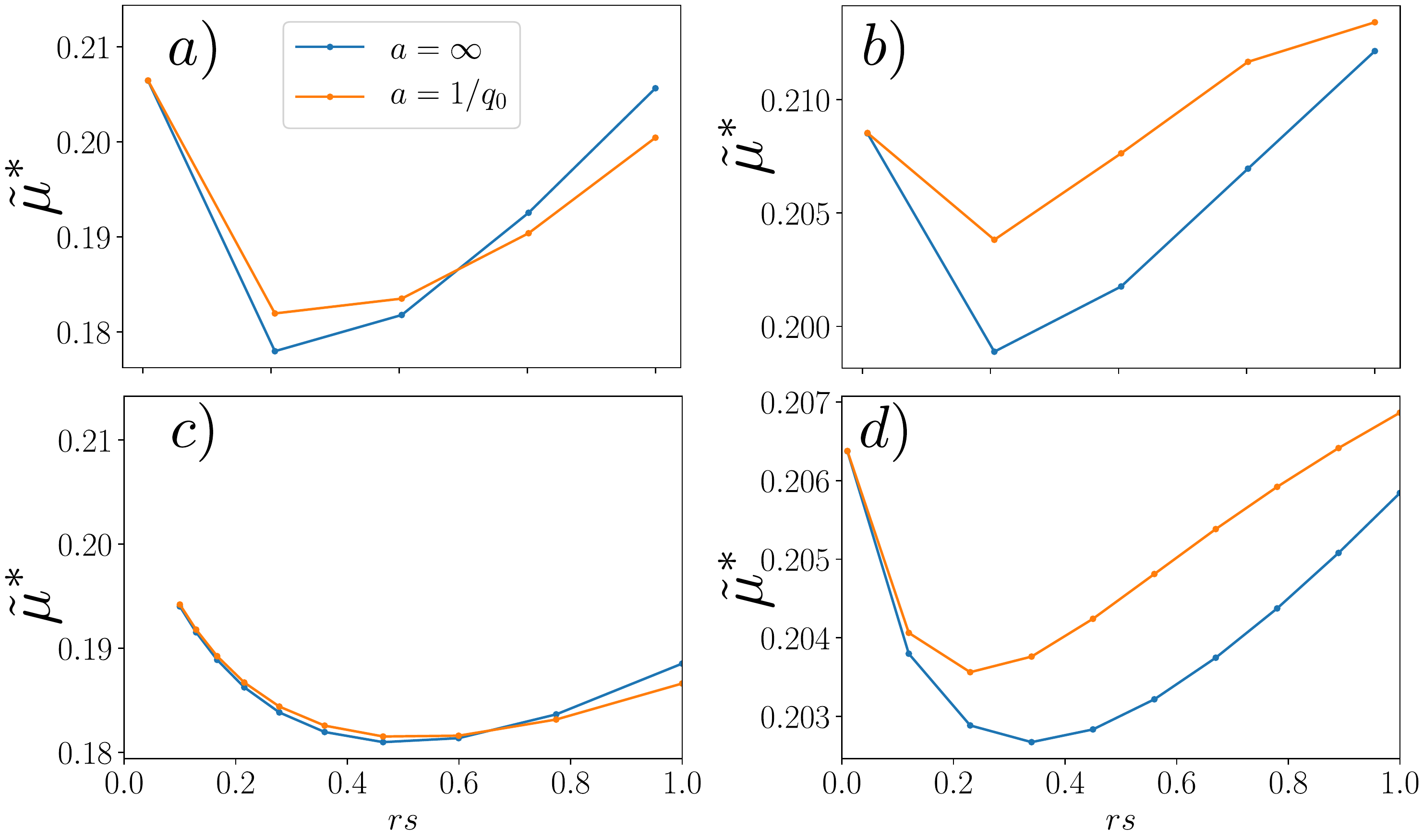}
\par\end{centering}
\caption{Superconducting coupling constant $\tilde{\mu}^{*}$ (eq.~\ref{eq:TcLambda})
extracted from solution of Migdal-Eliashberg equations.
a) $s$-wave channel, b) $d$-wave channel. The  toy-model results in the $s$- and $d$-wave channels are shown in (c, d). We take the following
parameters for the additional coupling constant projected onto the corresponding angular momentum channel ${\cal V}=-\lambda\omega_0^{2}/(\omega_0^{2}+\Omega_{m}^{2}$):
$\lambda=0.5,$ $\omega_0=E_{F}/10$. }

\label{Fig6}
\end{figure}

We now solve the gap equation Eq.~\eqref{eq:phi-1} including the additional attractive interaction Eq.~\eqref{eq:Vkkmod}. The result
of full numerical solution is shown in Fig.~\ref{Fig6}~(a, b). We find an insignificant
enhancement of superconductivity in the $s$-wave channel for $r_{s}<0.5$.
In contrast, in the $d$-wave channel we find enhancement in a broad range of parameters as was expected from our toy model.
 As can be seen from Fig.~\ref{Fig6}~(c, d) the toy-model results are in good qualitative agreement with the  Migdal-Eliasberg equation. In the following section we provide an example of modifying the hyperbolic metamaterial structure to achieve a plasmon dispersion that both enhances the superconductivity more efficiently  in the $d$-wave channel and achieves a noticeable enhancement in the conventional $s$-wave channel.

\subsection{Engineered plasmon-assisted enhancement\label{subsec:Engineering-strong-plasmon}}

\begin{figure}
\begin{centering}
\includegraphics[scale=0.3]{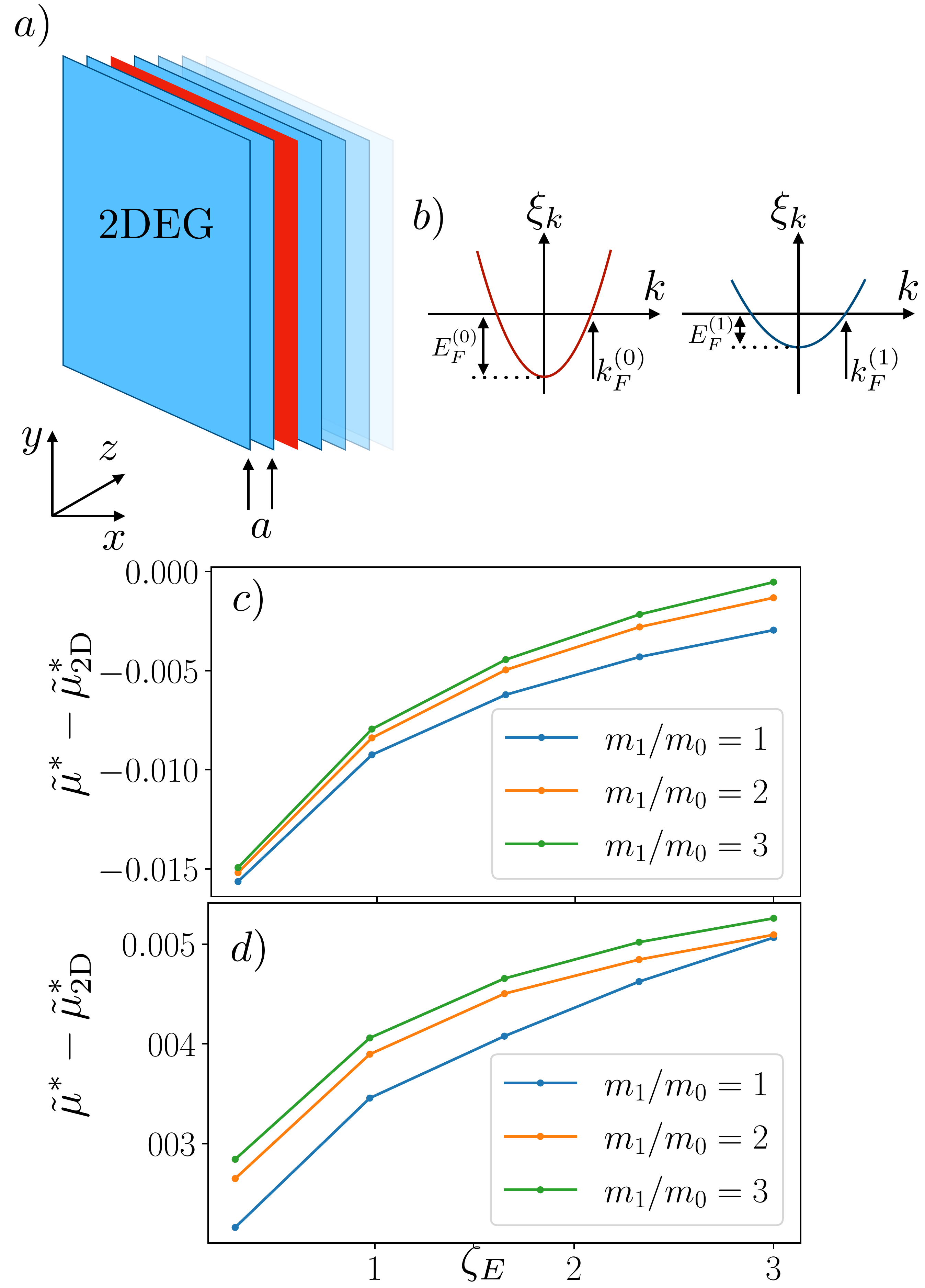}
\par\end{centering}
\caption{Engineered enhancement of superconductivity in a layered electron
gas. a) Setup of an infinitely large layered electron gas with inter-layer
distance $a$ the layer having distinct properties is shown in red.
b) Electron properties on different layers: we assume massive dispersions
but with the same Fermi momentum. Coupling constant 
$\tilde{\mu}^{*}$ in (c) $s$-wave and (d) $d$-wave (orange) channels as function
of $\zeta_{E}$ for different electron mass ratios. The inter-layer distance and Wigner-Seitz parameters are respectively chosen to be $a=1/q_{0}$, $r_{s}=1$.  The additional pairing parameters are $\lambda=1$, $\omega_0=0.1 E_F$}

\label{FIG10}
\end{figure}

In this section, we show how the superconducting transition temperature
can be enhanced by varying the electronic properties on one of the
layers ($i=0$) as shown in Fig.~\ref{FIG10}~(a, b). Guided by
the qualitative analysis of Sec.~\ref{sec:Discussion}, we seek to
reduce the screening energy range $\varepsilon_{\text{sc}}$ on the
layer shown in red, which we label as $i=0$. This is expected to
have a positive effect on the coupling strength based on the transition
temperature estimate in Eqs.~\eqref{eq:Tc} and~\eqref{eq:TcLambda}.
As a simple proof-of-principle example, we show that this can occur
if the target $i=0$ layer is distinct from the other layers. Experimentally
this may correspond to inserting a layer of a different material with
a higher standalone Fermi energy than that in the other layers (it
may also occur by targeted gating/doping or naturally on a boundary
layer). We do not specify a particular experimental realization, but
point out that it seems to be a reasonable setup, as Fermi energies
in different metals may vary by almost an order of magnitude \citep{AM76}.
For a quantitative description of the realistic metamaterial structures
of this type, we may need to consider complications associated with
a redistribution of charges, induced surface potentials in different
layers due to tunneling, Volta barrier, etc. The form and shape of
Fermi surfaces in different materials may also play a role.

We however disregard these possible complications and consider the
simplified toy model instead, where both materials have a circular
electronic dispersion, $\xi_{k}^{i}=k^{2}/2m_{i}-E_{F}^{(i)}$, with
the the electron gas on the zeroth layer having a distinct Fermi energy
$E_{F}^{(0)}$ and Fermi momentum $k_{F}^{(0)}$, while the other
layers are characterized by $E_{F}^{(1)}$ and $k_{F}^{(1)}$ as shown
in Fig.~\ref{FIG10}~(b). 
 Denoting the bare
Coulomb interaction matrix as ${\cal \widetilde{V}}_{{\bf q}}^{(i,j)}$,
 the complete RPA-renormalized  interaction $\check{{\cal V}}_{{\bf q}}^{(i,j)}$
can be written in the following dimensionless matrix form (defined
in the layer space parameterized by $i$ and $j$):

\begin{equation}
(\check{{\cal V}}_{{\bf q}}^{-1})^{(i,j)}=({\cal \widetilde{V}}_{{\bf q}}^{-1})^{(i,j)}+\delta_{i,j}q_{1}\Pi_{{\bf q}\zeta_{K}}\left(i\Omega_{n}\zeta_{E}\right)+\delta_{i,0}\delta_{j,0}\varXi_{{\bf q}},\label{eq:Vhat}
\end{equation}
where $\varXi_{{\bf q}}=q_{0}\Pi_{{\bf q}}\left(i\Omega_{n}\right)-q_{1}\Pi_{{\bf q}\zeta_{K}}\left(i\Omega_{n}\zeta_{E}\right)$,
$\zeta_{E}=E_{F}^{\left(0\right)}/E_{F}^{\left(1\right)}$, $\zeta_{K}=k_{F}^{\left(0\right)}/k_{F}^{\left(1\right)}$
and $q_{0}$, $q_{1}$ are the Thomas-Fermi
vectors of the $i=0$ and the rest layers respectively. 
The first two terms to the righthand side of Eq. \eqref{eq:Vhat}
represent the inverse of the Coulomb interaction matrix in an infinite
system of identical layers henceforth denoted as $({\cal V}_{{\bf q}}^{-1})^{(i,j)}\equiv({\cal \widetilde{V}}_{{\bf q}}^{-1})^{(i,j)}+\delta_{i,j}q^{(1)}\Pi_{{\bf q}\zeta_{K}}\left(i\Omega_{n}\zeta_{E}\right)$.
The matrix inversion in Eq.~\eqref{eq:Vhat} can be performed using
the Sherman-Morrison formula \citep{SM50}:

\begin{align}
\check{{\cal V}}_{{\bf q}}^{(i,j)} & ={\cal V}_{{\bf q}}^{(i,j)}-\varXi_{{\bf q}}\frac{{\cal V}_{{\bf q}}^{\left(i,0\right)}{\cal V}_{{\bf q}}^{\left(0,j\right)}}{1+{\cal V}_{{\bf q}}^{\left(0,0\right)}\varXi_{{\bf q}}},\label{eq:Vhat-1}
\end{align}
We thus expressed the interaction Eq.~\eqref{eq:Vhat} in terms
of the Coulomb propagator of a translationally-invariant system. The
latter can be straightforwardly found  in Fourier space. In particular, for the interaction on $i=0$ layer
we get:

\begin{align}
\check{{\cal V}}_{{\bf q}}^{\left(0,0\right)} & =\frac{{\cal V}_{{\bf q}}^{\left(0,0\right)}}{1+{\cal V}_{{\bf q}}^{\left(0,0\right)}\varXi_{{\bf q}}}.\label{eq:VV}
\end{align}
We now calculate the coupling constant induced by this interaction.
We employ the following parametrization $\zeta_{K}=\sqrt{\zeta_{E}m_{1}/m_{0}}$
and vary only the $\zeta_{E}$ parameter for different electron mass
ratios. Larger values $\zeta_{E}$ correspond to the lower value of the Fermi and screening energies $\varepsilon_{\text{sc}}$
of $i\neq0$ layers. Based on the qualitative analysis in Sec.~\ref{sec:Discussion} we
thus expect to increase the coupling strength on $i=0$ layer. Numerically obtained coupling constant $\tilde{\mu}^{*}$ is shown
in Fig.~\ref{FIG10}~(c, d) for pairing in $s$- and $d$-wave channels
respectively. In case of the $d$-wave channel the additional enhancement
of the coupling constant is found for all values $\zeta_{E}>1$. 
 In the $s$-wave channel at low values of  $\zeta_E$ we find that the layering is decreasing $\tilde{\mu}^{*}$.
We also find that higher values of  $\zeta_E$ tend to increase the coupling strength in accord with the qualitative discussion above. 
As a result we find that there is a threshold value of $\zeta_E$ above which the engineered layered structure enhances the electron pairing. 
In particular, for the electron mass
ratio $m_{1}/m_{0}=3$ the threshold is  $\zeta_{E}\approx3$. At smaller electron mass rations the enhancement
is expected to occur at larger values of $\zeta_{E}$.

In conclusion, in this section we studied the plasmon-assisted pairing in layered metals. We found a significant enhancement in the $d$-wave channel in the case of a perfectly regular metallic array. We also provided a way to engineer the plasmon environment in a layered metal such as to enhance Cooper pairing of electrons in both $s$- and $d$-wave channels.

\section{Conclusions}

The interplay between Coulomb interactions and superconductivity has been studied since the early days of superconductivity research. This includes numerous studies of the effect of plasmons on phonon-mediated superconductivity and also scenarios of plasmon-mediated superconductivity. It is therefore somewhat surprising that a standard Migdal-Eliashberg  treatment of plasmons in layered materials, of obvious interest to the cuprates for example, has not been performed until this work. Here, we systematically studied various effects of plasmons in layered structures  on superconductivity using both full Migdal-Eliashberg theory and qualitative arguments, which allowed us to develop useful intuition. 

While we find neither an enhancement of $s$-wave pairing nor a plasmon-only pairing in hyperbolic structures, we do find a surprising effect of a sizable enhancement of $d$-wave pairing there. What it implies is that given a $d$-wave intrinsic superconductor (e.g., thin superconducting film), layering such films would lead to a significant enhancement of $T_c$. Furthermore, we point out that the ``simple'' layering is not the only possible way to affect the plasmon physics, and show that realistic paths exist to bootstrap both conventional $s$-wave and unconventional superconductivity to higher temperatures by engineering more favorable plasmon dispersions. 

Apart from a possible connection to the layered oxide superconductors, there are additional arguments for why such a study is of interest and timely. Namely, there has been much interest~\citep{CRA19,ARC19,SCJ19,SRR18,TEN19,KCX21,CFL20,SG18} recently in engineering electromagnetic environment \citep{RGS18, FR18} in metamaterials and using cavities to achieve an enhancement of superconductivity. Of particular note are a series of works by Smolyaninova et al.~\citep{SJZ16,SYZ14}, where a clear and significant enhancement of both transition temperature and critical field has been observed in fabricated aluminum/aluminum oxide layered structures. It was speculated that this could happen because of the anomalous hyperbolic plasmon dispersion, which may mediate superconductivity. The present work seems to rule out this scenario. This negative result however points towards phonons as a ``culprit.'' In particular, related hyperbolic phononic modes are of interest and will be discussed in a separate study. 

\centerline{\bf Acknowledgements}
The authors are grateful to Igor Smolyaninov and Sankar Das Sarma for helpful discussions.
This work was supported by the National Science Foundation under Grant No. DMR-2037158, the U.S. Army Research Office under Contract No. W911NF1310172, and the Simons Foundation.

\bibliographystyle{apsrev4-1}
\bibliography{bibl}

\appendix

\section{Analytical solution\label{sec:Analytical-solutions}}

In this section we provide an exact solution of the gap equation in the case of a simplistic model, where the energy-dependence of the interaction is separable and is a product of two step-functions -- a ``square-well'' model.

\subsubsection{``Square-well'' model \label{subsec:Square-well-model}}

In order to get an analytical estimate of the superconducting transition
temperature we assume the following separable form of the interaction in Eq.~(\ref{eq:Vtildetilde}):
$v_{m,n}\approx v_{n}v_{m}$ with $v_{n}=\theta(\varepsilon_{\text{sc}}^{2}-\varepsilon_{n}^{2})$.
The gap equation becomes
\begin{align}
  \phi^{\left(i,j\right)}\left(i\varepsilon_{n}\right)=
  -\frac{\mu}{\beta}\sum_{m}\left(1-\sigma v_{n}v_{m}\right)f_{m}\phi^{(l)}\left(i\varepsilon_{m}\right),\label{eq:Eq_GS-1}
\end{align}
where we denoted $f_{m}=\int_{-E_{F}}^{E_{F}}d\xi_{k'}/(\varepsilon_{m}^{2}+\xi_{k'}^{2})$.
We note that since we work with dimensionless units, the 
Fermi energy is  $E_{F}=1/4$. Due to the separability
of the interaction, we define $A=\sum_{n}v_{n}f_{n}\phi^{\left(i,j\right)}\left(i\varepsilon_{n}\right)$
and $B=\sum_{n}\left(1-v_{n}\right)f_{n}\phi^{\left(i,j\right)}\left(i\varepsilon_{n}\right)$,
which form the closed set of equations:

\begin{align*}
A= & -\frac{\mu}{\beta}\left(A+B\right)\left(\sum f_{n}v_{n}\right)+\left(\sum f_{n}v_{n}\right)\frac{\mu\sigma}{\beta}A,\\
B= & -\frac{\mu}{\beta}\left(A+B\right)\sum_{n}\left(1-v_{n}\right)f_{n}.
\end{align*}
These equations lead to the self-consistency equation, which determines temperature condition.
\begin{align*}
1= & \frac{\mu}{\beta}\left(\sigma-\frac{1}{1+\frac{\mu}{\beta}\sum_{n}\left(1-v_{n}\right)f_{n}}\right)\sum f_{n}v_{n}
\end{align*}
We now use the following relations:
\begin{align*}
&T&\!\!\!\!\!\! \sum_{n}f_{n}v_{n} & \approx\log\frac{2e^{\gamma}\varepsilon_{\text{sc}}\beta}{\pi},\\
&T&\!\!\!\!\!\!\!\!\!\sum_{n}f_{n} & =\int_{0}^{E_{F}}d\xi_{k}\frac{\tanh\frac{\beta\xi_{k}}{2}}{\xi_{k}}
 \approx\log\frac{2e^{\gamma}E_{F}\beta}{\pi}.
\end{align*}
By substituting this form to the Migdal-Eliashberg equation Eq.~\eqref{eq:Eq_GS-1} and using (\ref{eq:Vtildetilde})
we readily find the transition temperature assuming $\varepsilon_{\text{sc}}\ll E_{F}$:

\begin{equation}
T_{c}=1.134\varepsilon_{\text{sc}}e^{-1/\mu_{\text{pl}}^{*}},\label{eq:T_c}
\end{equation}
where 
\begin{equation}
\mu_{\text{pl}}^{*}=\mu\left(\sigma-\frac{1}{1+\mu\log\frac{E_{F}}{\varepsilon_{\text{sc}}}}\right).\label{eq:mustar}
\end{equation}
Superconductivity is only present if $\mu^{*}>0$.
We observe that the Coulomb repulsion is strongly suppressed at the energy scales of order $\varepsilon_{\text{sc}}\ll E_{F}$ \citep{P92}. Although
the realistic interaction potential (see main text)  is more complex than the one used in
this Appendix, the qualitative arguments behind the result here --  Eqs.~(\ref{eq:T_c}) and~(\ref{eq:mustar}) --
are universal. Specifically, the  transition temperature is primarily determined by two  dimensionless
parameters: $\sigma$ (effective screening strength) and $\varepsilon_{\text{sc}}/E_{F}$ (effective screening energy range).

\section{Pseudopotential method}

In this section we provide details on the pseudopotential method, which we use to  determine the effective pairing strength. The derivation follows Ref.~\citep{GS84}. We start with the gap equation Eq~\eqref{eq:Eq_GS}:
\begin{align}
  \phi^{(l)}\left(i\varepsilon_{n}\right)=\nonumber 
  -T\sum_{m}{\cal V}\left(i\varepsilon_{n}-i\varepsilon_{m}\right)f_{m}\phi^{(l)}\left(i\varepsilon_{m}\right),\label{eq:Eq_GS-1-1}
\end{align}
where $f_{m}=\int_{-E_{F}}^{E_{F}}d\xi_{k'}\frac{1}{\varepsilon_{m}^{2}+\xi_{k'}^{2}}$.
We now introduce a low-energy frequency cut-off function $v_{n}=\theta\left(\varepsilon_{\text{c}}^{2}-\varepsilon_{n}^{2}\right)$
with the assumption $\varepsilon_{\text{c}}\ll\varepsilon_{\text{sc}}\ll E_{F}$.

\begin{align*}
  \phi^{(l)}\left(i\varepsilon_{n}\right)=
 & -T\sum_{m}{\cal V}\left(i\varepsilon_{n}-i\varepsilon_{m}\right)f_{m}v_{m}\phi^{(l)}\left(i\varepsilon_{m}\right)\\
 & -T\sum_{m}{\cal V}\left(i\varepsilon_{n}-i\varepsilon_{m}\right)f_{m}\left(1-v_{m}\right)\phi^{(l)}\left(i\varepsilon_{m}\right).
\end{align*}
With this, we now integrate out the high-energy modes and find,
\begin{equation}
\phi^{(l)}\left(i\varepsilon_{n}\right)=-T\sum_{m}{\cal V}_{n,m}^{\text{eff}}f_{m}v_{m}\phi^{(l)}\left(i\varepsilon_{m}\right)\label{eq:Eq_Pseudo}
\end{equation}
with ${\cal V}^{\text{eff}}$ obeying the equation,
\begin{align}
 & {\cal V}_{n,m}^{\text{eff}}=\label{eq:Veff}\\
 & {\cal V}\left(i\varepsilon_{n}-i\varepsilon_{m}\right)-T\sum_{n'}{\cal V}\left(i\varepsilon_{n}-i\varepsilon_{n'}\right)\left(1-v_{n'}\right){\cal V}_{n',m}^{\text{eff}}.\nonumber 
\end{align}
If the cut-off energy is low enough, we can safely assume $\phi_{n}^{(l)}\approx\phi^{(l)}\left(i\varepsilon_{n=0}\right)$
and ${\cal V}_{n,m}^{\text{eff}}\approx{\cal V}_{0,0}^{\text{eff}}$.
The critical temperature is then

\[
T\approx\frac{2e^{\gamma}\varepsilon_{\text{co}}}{\pi}e^{\frac{1}{V_{0,0}^{\text{eff}}}}\approx1.134\varepsilon_{\text{co}}e^{\frac{1}{V_{0,0}^{\text{eff}}}}
\]
We thus identify the term $(-V_{0,0}^{\text{eff}})$ as the Coulomb
pseudopotential at   energy $\varepsilon_{\text{c}}$. In this approach, the problem of finding the critical temperature reduces to solving Eq.~\eqref{eq:Veff}. The latter represents a system of linear equations, which can be solved numerically. An estimate can 
be obtained for the separable potential in analogy with the previous Appendix~A.

\section{Kirzhnits-Maksimov-Khomskii theory \label{subsec:Kirzhnits-Maksimov-Khomskii-theo}}

\begin{figure}
\begin{centering}
\includegraphics[scale=0.3]{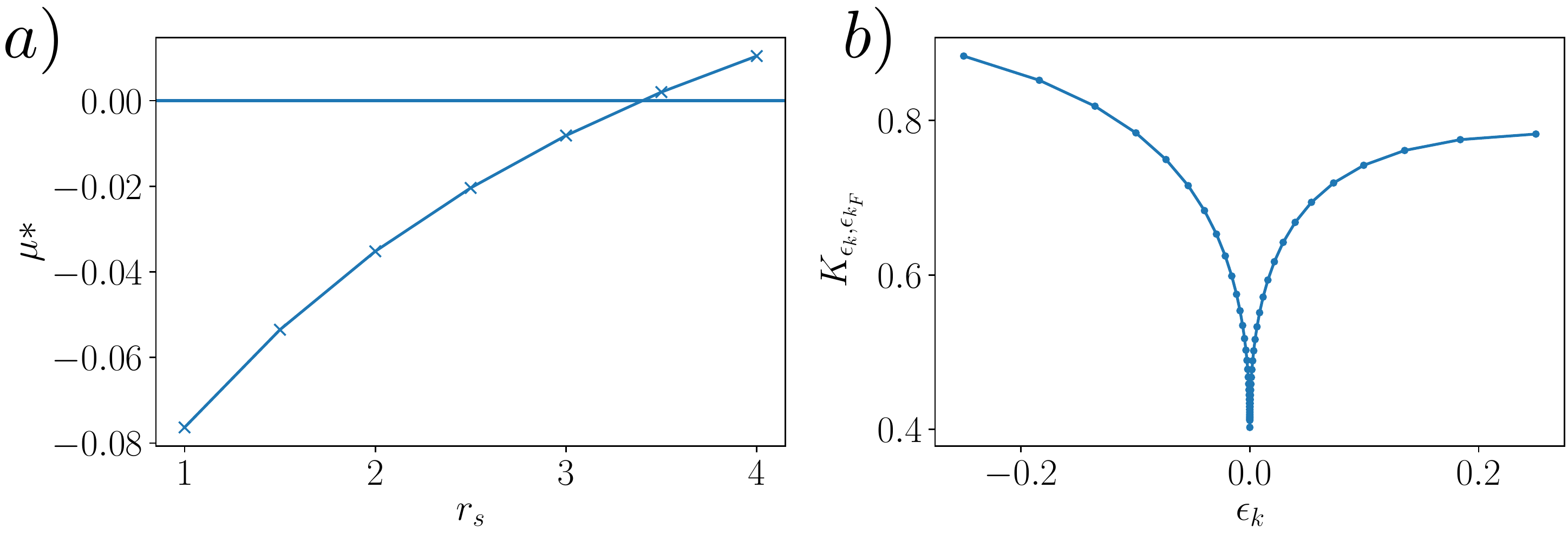} 
\par\end{centering}
\caption{a) Coulomb pseudopotential for $a\rightarrow\infty$. b) effective
interaction $K$ expressed in energy Eq.~(\ref{eq:K_pk}) for $r_{s}=2.5$. }

\label{KMK} 
\end{figure}

Here we briefly discuss the commonly-used approach by Kirzhnits, Maksimov, and Khomski (KMK)  to plasmon-induced superconductivity based on Ref.~\citep{KMK73}. This theory deals with the gap equation in the real frequency domain as opposed
to the Matsubara formalism in the conventional Eliashberg formulation presented in the main text.  The gap equation in the KMK approach is of the BCS form:

\begin{equation}
\Delta_{p}=-\int d\xi_{k}K_{p,k}\frac{\tanh\frac{\beta\xi_{k}}{2}}{2\xi_{k}}\Delta_{k},\label{eq:KMK}
\end{equation}
where the gap is defined as $\Delta_{p}\equiv2\epsilon_{p}\int_{0}^{\infty}dxf\left(x,p\right)$
and $f$ is the analytical continuation of the pair propagator $F_{{\bf q},n}=-\int e^{i\varepsilon_{n}\tau}\langle\hat{\psi}_{{\bf q}}^{\left(i\right)}\left(\tau\right)\hat{\psi}_{-{\bf q}}^{\left(i\right)}\left(0\right)\rangle$.
The kernel of the KMK gap equation (\ref{eq:KMK})  is given
by
\begin{equation}
K_{p,k}=V_{p,k}^{\left(0\right)}-2\int_{0}^{\infty}d\varepsilon\frac{\text{Im}V_{p,k}^{R}\left(\varepsilon\right)}{\varepsilon+\left|\epsilon_{p}\right|+\left|\xi_{k}\right|},\label{eq:K_pk}
\end{equation}
where $V_{p,k}^{R}$ is the analytically continued interaction Eq.~(\ref{eq:Vii}).
KMK have derived Eq. (\ref{eq:KMK}) by performing the analytical continuation
in the Eliashberg gap equation (\ref{eq:phi-1}) and  neglecting several
terms that are not singular at the transition point. However, these terms are still be important for correct determination
of the transition temperature, as was argued in Refs.~\citep{SS83,KA80}. The transition temperature
in Eq. (\ref{eq:KMK}) can be estimated analytically  assuming the
separability of the interaction kernel $K$ analogously to Sec.~\ref{sec:Analytical-solutions}.
We solve Eq.~(\ref{eq:KMK}) numerically \citet{T78}. The typical pairing strength is shown in Fig.~\ref{KMK}. We find that the KMK approximation severely underestimates the pairing strength compared to the  Migdal-Eliashberg equations. We therefore conclude that the KMK approach is quantitatively unreliable.

\section{Effects of disorder in layer positions}

In this section we study the effect of disorder in layer positions
on the superconducting transition. For that we take the bare Coulomb
interaction matrix as $\widetilde{{\cal V}}_{{\bf q}}^{(i,j)}=e^{-a_{i,j}q\left|i-j\right|}/q$,
where $a_{i,j}$ is the set of random interlayer spacings. In the
following we will treat $\widetilde{{\cal V}}_{{\bf q}}^{(i,j)}$ as a
matrix in the layer-index space as in Sec.~\ref{subsec:Engineering-strong-plasmon}.
For the inter-layer spacings we assume $a_{i,j}=a_{i,j}^{(0)}+a_{i,j}^{\delta},$
where $a_{i,j}^{\delta}$ is a uniformly distributed random variable
$a_{i,j}^{\delta}\in[0,\delta]$ and $a_{i,j}^{(0)}$ is the minimal
inter-layer distance. Adopting the formalism of Sec.~\ref{subsec:Engineering-strong-plasmon},
we write the interaction matrix in the absence of translational symmetry
along $z$ axis as 
\begin{equation}
({\cal V}_{{\bf q}}^{-1})^{(i,j)}=\left({\cal \widetilde{V}}_{{\bf q}}^{-1}\right)^{(i,j)}+\delta_{i,j}q_{0}\Pi_{{\bf q}}\left(i\Omega_{n}\right).\label{eq:Vij_disorder}
\end{equation}
Below we find the transition temperature corresponding to Eliashberg
equations with interaction (\ref{eq:Vij_disorder}).

\subsubsection{Plasmon localization}

\begin{figure}
\begin{centering}
\includegraphics[scale=0.3]{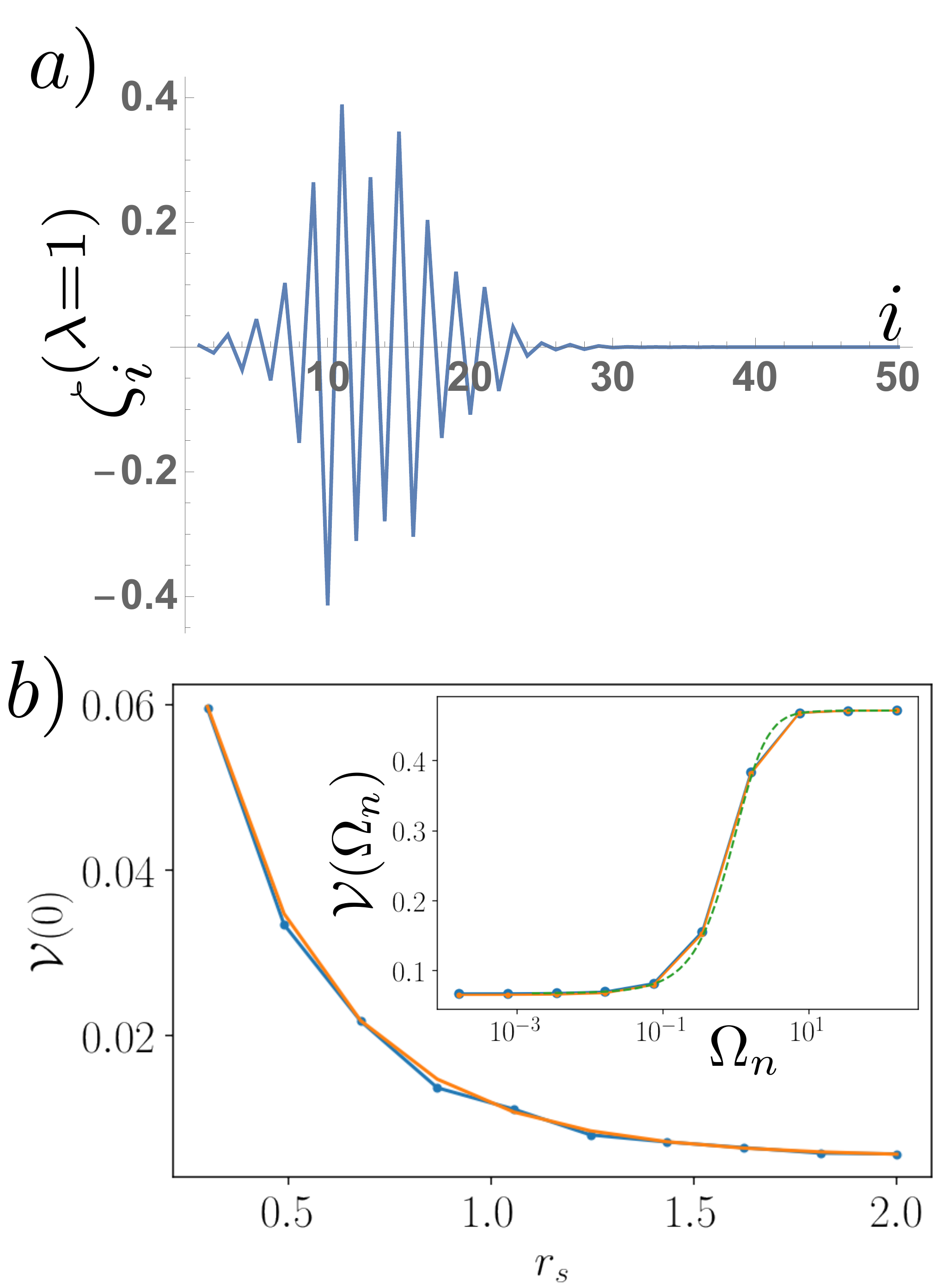} 
\par\end{centering}
\caption{Anderson localization of plasmons. a) Typical plasmon distribution
for the lowest eigenvalue $\lambda=1$ (Eq.~\ref{eq:Eigenvalue}),
b) Typical static interaction ${\cal V}\left(0\right)$ as function
of the electron gas Wigner-Seitz density parameter $r_{s}$ for 8
layers: the random layer spacing $a_{i,j}=a_{i,j}^{(0)}+a_{i,j}^{\delta}$
shown in blue with the average spacing $a_{i,j}^{\left(0\right)}=q_{0}^{-1}$
and $\delta=q_{0}^{-1}$, regular array result shown in orange. Inset
shows the interaction as function of the imaginary frequency $\Omega_{n}$.
The infinite layer number is shown in green color for comparison. }

\label{FIG_AL} 
\end{figure}

Plasmon modes in a disordered layered gas correspond to the
eigenvalues of the matrix (\ref{eq:Vij_disorder}). We denote them
as $\left\{ \lambda\right\} $ and the corresponding eigenvectors
as $\{\zeta_{j}^{\left(\lambda\right)}\}$: 
\begin{equation}
\sum_{j}{\cal V}_{{\bf q}}^{(i,j)}(i\Omega_{n})\zeta_{j}^{(\lambda)}=\epsilon_{\lambda}\zeta_{i}^{(\lambda)}.\label{eq:Eigenvalue}
\end{equation}
Here, the eigenvalues are functions of both the in-plane momentum
${\bf q}$ and the frequency $i\Omega_{n}$: $\epsilon_{\lambda}=\epsilon_{\lambda}\left(i\Omega_{n}\right)$.
The typical eigenvector of Eq.~\eqref{eq:Eigenvalue} is shown in
Fig.~\ref{FIG_AL}~(a), where performed the analytic continuation
to real frequencies $i\Omega_{n}\rightarrow\omega+i0^{+}$. We observe
Anderson localisation \cite{SKP86} of plasmons in the apical direction
\citep{A58} for $N=50$ layers. In Fig.~\ref{FIG_AL}~(b) we study
the effective interaction between layers close to the center of the
sample. More precisely, we consider the approach of Sec.~\eqref{sec:Discussion}
for the disorder interaction given in Eq.~\eqref{eq:Vij_disorder}.
As can be seen in Fig.~\ref{FIG_AL}~(b) the interaction is barely
affected by the disorder. We therefore conclude that disorder does
not modify the critical temperature. 

\section{Finite thickness of layers}

\begin{figure}
\begin{centering}
\includegraphics[scale=0.4]{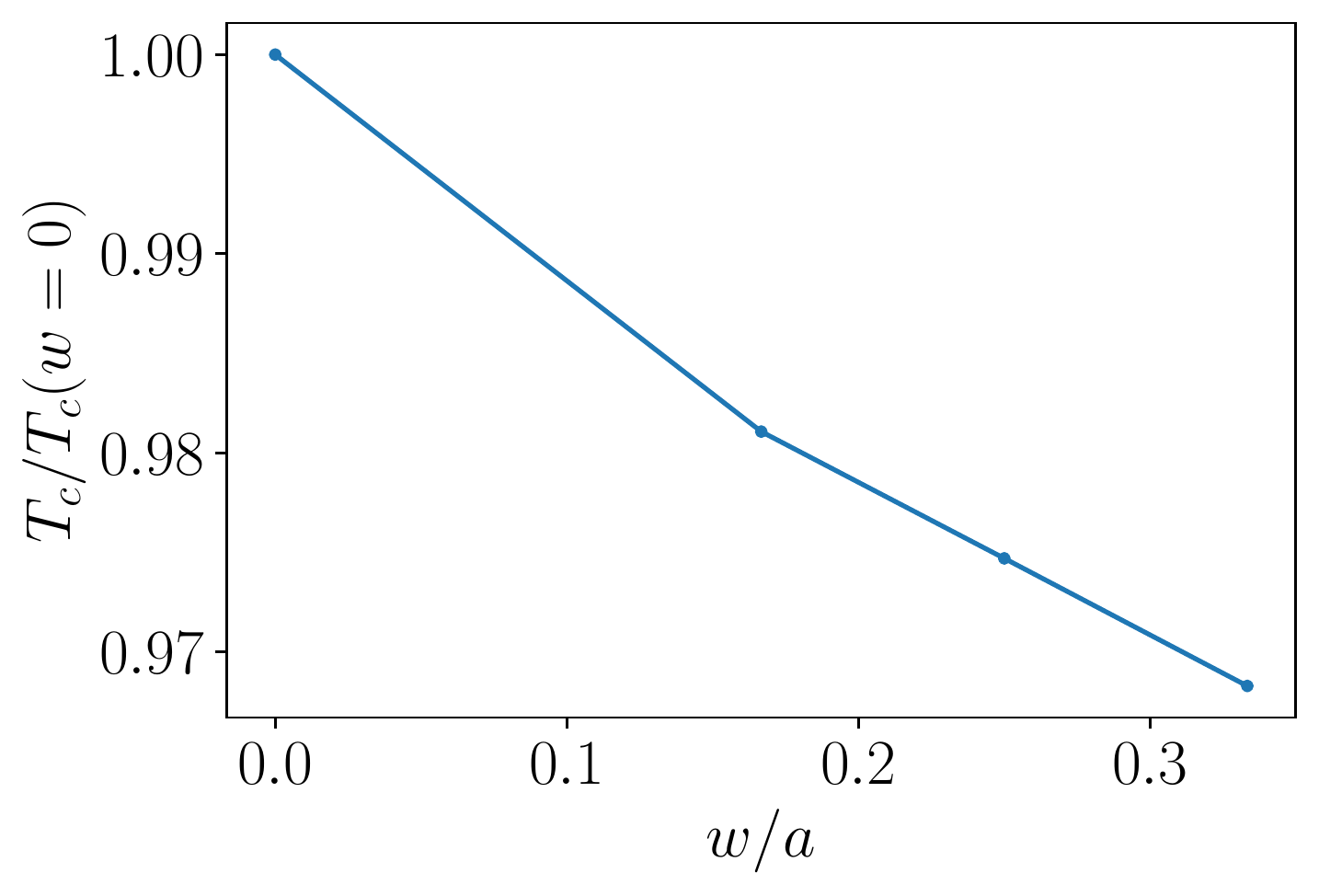} 
\par\end{centering}
\caption{Superconducting transition temperature as function of layer thickness.
Simulation parameters are chosen as $a=1/q_{0}$, $r_{s}=1$, $\lambda=1$.}

\label{Fig15} 
\end{figure}

In this section we consider the setup with the finite-thickness
layers \cite{BMS20,BS17}. In this section we follow the approach
\cite{S84}. For a given thickness $w$ of each layer we now need
to keep the quantization of electron wavefuntion along the $z$ direction.
The generalization of electron polarization operator can be written
as:

\begin{align*}
\Pi_{{\bf q}}^{\left(w\right)}\left(i\Omega_{m},z,z'\right) & =\frac{T}{A}\sum_{{\bf k},n}\sum_{l,l'}\frac{\zeta_{l}\left(z\right)\zeta_{l}\left(z'\right)}{i\varepsilon_{n}-\epsilon_{{\bf k}}-\frac{k_{l}^{2}}{2m}}\\
 & \times\frac{\zeta_{l'}\left(z\right)\zeta_{l'}\left(z'\right)}{i\varepsilon_{n}+i\Omega_{m}-\epsilon_{{\bf k}+{\bf q}}-\frac{k_{l'}^{2}}{2m}},\\
\end{align*}
where $\zeta_{l}\left(z\right)\equiv\sqrt{2/w}\sin(k_{l}z)$ denotes
the $l$-th subband mode profile of the electron gas and $k_{l}=\pi l/w$
with $l\in\mathbb{Z}$. Here we restrict to the case when layers are
sufficiently thin and restrict our consideration to the lowest-energy
sub-band:

\begin{equation}
\Pi_{q}^{\left(w\right)}\left(z,z'\right)\approx\zeta_{1}^{2}\left(z\right)\zeta_{1}^{2}\left(z'\right)\Pi_{q},\label{eq:Pi^w}
\end{equation}
where $\Pi_{q}$ is the usual two-dimensional polarization operator.
It is now convenient to project the Coulomb interaction onto the lowest
sub-band wavefunction. In particular, for the bare interaction we have:

\begin{align*}
{\cal \widetilde{V}}_{{\bf q}}^{\left(i,j\right)}\left(w\right) & \equiv\frac{1}{q}\int dzdz'e^{-q\left|z-z'\right|}\zeta_{1}^{2}\left(z+ai\right)\zeta_{1}^{2}\left(z'+aj\right).
\end{align*}
This integral can be taken analytically. In particular for $i\neq j$
we find:

\begin{equation}
{\cal \widetilde{V}}_{{\bf q}}^{\left(i\neq j\right)}\left(w\right)=\widetilde{{\cal V}}_{{\bf q}}^{\left(i\neq j\right)}\left(w=0\right)\times\frac{16\pi^{4}e^{-wq}\left(e^{wq}-1\right)^{2}}{w^{2}q^{2}\left(w^{2}q^{2}+4\pi^{2}\right)^{2}}.\label{eq:ViNEQj}
\end{equation}
The term $i=j$ is more cumbersome and we will not reproduce it here.
We now transform into Fourier space with respect to the layer index
and get:

\begin{align*}
\widetilde{{\cal V}}_{{\bf q},q_{z}}\left(w\right) & =\sum_{j\neq0}e^{-iq_{z}j}\widetilde{{\cal V}}_{{\bf q}}^{\left(j,0\right)}\left(w\right)+\widetilde{{\cal V}}_{{\bf q}}^{\left(0,0\right)}\left(w\right)\\
 & =\frac{1}{q}\frac{\sinh aq}{\cosh aq-\cos q_{z}}\frac{16\pi^{4}e^{-wq}\left(e^{wq}-1\right)^{2}}{w^{2}q^{2}\left(w^{2}q^{2}+4\pi^{2}\right)^{2}}\\
 & +\widetilde{{\cal V}}_{{\bf q}}^{\left(0,0\right)}\left(w\right)-\frac{1}{q}\times\frac{16\pi^{4}e^{-wq}\left(e^{wq}-1\right)^{2}}{w^{2}q^{2}\left(w^{2}q^{2}+4\pi^{2}\right)^{2}}
\end{align*}
We this we can now repeat the same procedure of RPA renormalization
of interaction.

\[
{\cal V}_{{\bf q},q_{z}}\left(i\Omega_{m},w\right)=\frac{1}{\widetilde{{\cal V}}_{{\bf q},q_{z}}^{-1}(w)+q_{0}\Pi_{{\bf q}}\left(i\Omega_{m}\right)}.
\]
As shown in Fig.~\ref{Fig15} the effect of layer thickness on the
transition temperature is insignificant under our approximations.

\end{document}